\documentclass[conference]{IEEEtran}
\usepackage[
backend=biber,
style=numeric,
sorting=none
]{biblatex}
\usepackage{tikz}
\usetikzlibrary{shapes.geometric, arrows}

\tikzstyle{startstop} = [rectangle, rounded corners, minimum width=2.5cm, minimum height=1cm, text centered, draw=black, fill=red!30]
\tikzstyle{process} = [rectangle, minimum width=2.5cm, minimum height=1cm, text centered, draw=black, fill=orange!30]
\tikzstyle{intermediate} = [rectangle, minimum width=2.5cm, minimum height=1cm, text centered, draw=black, fill=green!30]
\tikzstyle{arrow} = [thick,->,>=stealth]

\usepackage{listings}
\usepackage{url}
\usepackage{amsmath,amssymb,amsfonts}
\usepackage{algorithmic}
\usepackage{graphicx}
\usepackage{textcomp}
\usepackage{xcolor}
\def\BibTeX{{\rm B\kern-.05em{\sc i\kern-.025em b}\kern-.08em
    T\kern-.1667em\lower.7ex\hbox{E}\kern-.125emX}}
\begin{document}

\title{FuzzDistill: Intelligent Fuzzing Target Selection using Compile-Time Analysis and Machine Learning
}

\author{\IEEEauthorblockN{Saket Upadhyay}
\textit{Ph.D. Student}
\IEEEauthorblockA{\textit{Dept. of Computer Science} \\
\textit{University of Virginia}\\
saket@virginia.edu}
}

\maketitle

\begin{abstract}
Fuzz testing is a fundamental technique employed to identify vulnerabilities within software systems. However, the process can be protracted and resource-intensive, especially when confronted with extensive codebases. In this work, I present FuzzDistill, an approach that harnesses compile-time data and machine learning to refine fuzzing targets. By analyzing compile-time information, such as function call graphs' features, loop information, and memory operations, FuzzDistill identifies high-priority areas of the codebase that are more probable to contain vulnerabilities. I demonstrate the efficacy of my approach through experiments conducted on real-world software, demonstrating substantial reductions in testing time.
\end{abstract}


\section{Introduction}
Fuzz testing is a critical technique for identifying vulnerabilities in software by subjecting programs to random or semi-random inputs. While effective, traditional fuzzing methods often struggle with efficiency due to the vast codebases and complex behaviors of modern software. As a result, large portions of the code are left unexplored, and significant vulnerabilities can go undetected.

Directed fuzzing has emerged as a solution to address these limitations by focusing testing efforts on areas of the code most likely to contain vulnerabilities. However, most of the existing approaches typically rely on runtime feedback, often overlooking valuable compile-time information that could offer deeper insights into a program’s structure and behavior.

Compile-time analysis offers valuable insights into code structure, data flow, and control flow, which can be instrumental in guiding fuzz testing endeavors. Simultaneously, machine learning and Neural Networks methodologies enable the development of sophisticated models capable of analyzing intricate patterns within extracted data.

This project explores the convergence of two distinct trends. FuzzDistill integrates compile-time data analysis with machine learning techniques to optimize target selection. By identifying high-priority areas of the codebase and directing testing efforts to these regions, FuzzDistill seeks to significantly reduce testing time and resource utilization.

\section{Background}
Fuzz testing is a widely employed technique for identifying vulnerabilities in software by providing invalid, unexpected, or random data inputs to a program in an attempt to elicit errors or crashes. While fuzzing has demonstrated efficacy in uncovering bugs, the extensive volume of code in contemporary software systems and the inherent unpredictability of the outcomes render it challenging to apply fuzz testing in a time-efficient and comprehensive manner.

Directed fuzzing seeks to enhance the efficiency of fuzz testing by directing the exploration of the codebase toward areas that are more likely to harbor security flaws. One promising approach to achieve this is through the utilization of compile-time data, which can provide valuable insights into the program’s structure. Features such as function call graphs, control flow information, and memory operations can unveil critical information regarding the software’s behavior and potential areas of vulnerability.\cite{zhu_regression_2021,weissberg_sok_2024,haller_dowsing_2013,osterlund_parmesan_2020}

Traditional methods are constrained by an over-reliance on high-level abstractions that overlook subtle, lower-level details of the code. 

Machine learning models, when applied to fuzz testing, can discern patterns and correlations in code that are predictive of areas likely to contain vulnerabilities.

Compile-time data that can offer a more granular perspective of the software’s structure prior to execution.

In this project, I explore an approach that combines compile-time data analysis with machine learning to optimize target selection in directed fuzzing. FuzzDistill refines the fuzzer’s focus by analyzing features such as function call graphs, loop structures, and memory operations, which provide valuable insights into the dynamic behavior of the program without the need for runtime information. By leveraging this compile-time data, FuzzDistill identifies areas of the code that are more likely to contain vulnerabilities, allowing for a more targeted and efficient fuzzing process.

This approach not only offers a promising enhancement to fuzzing efficiency but also serves as a proof of concept for integrating static analysis and machine learning in the context of security testing.

\section{Architecture}
FuzzDistill\cite{noauthor_saket-upadhyayfuzzdistillcc_nodate, noauthor_saket-upadhyayfuzzdistillml_nodate, noauthor_saket-upadhyayfuzzdistillweb_nodate} is made up of three components,
\begin{enumerate}
    \item FuzzDistillCC: Compiler back-end for feature extraction,
    \item FuzzDistillML: Model training component, and 
    \item FuzzDistillWeb: Prediction front-end
\end{enumerate}

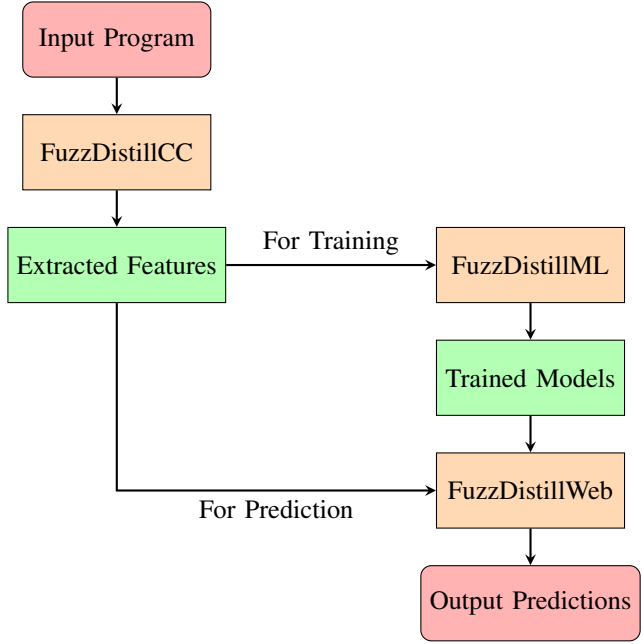
\begin{figure}[hbt!]
    \centering
    \begin{tikzpicture}[node distance=1.5cm and 2cm]

    
        \node (input) [startstop] {Input Program};
        \node (FuzzDistillCC) [process, below of=input] {FuzzDistillCC};
        \node (features) [intermediate, below of=FuzzDistillCC] {Extracted Features};
        \node (FuzzDistillML) [process, right of=features, xshift=4cm] {FuzzDistillML};
        \node (models) [intermediate, below of=FuzzDistillML] {Trained Models};
        \node (FuzzDistillWeb) [process, below of=models] {FuzzDistillWeb};
        \node (output) [startstop, below of=FuzzDistillWeb] {Output Predictions};
        
  
        \draw [arrow] (input) -- (FuzzDistillCC);
        \draw [arrow] (FuzzDistillCC) -- (features);
        \draw [arrow] (features) -- node[above, pos=0.5] {For Training} (FuzzDistillML); 
        \draw [arrow] (features) |- node[below, pos=0.75] {For Prediction} (FuzzDistillWeb); 
        \draw [arrow] (FuzzDistillML) -- (models);
        \draw [arrow] (models) -- (FuzzDistillWeb);
        \draw [arrow] (FuzzDistillWeb) -- (output);

    \end{tikzpicture}
    \caption{Workflow of FuzzDistill}
    \label{fig:fuzzdistill_flowchart}
\end{figure}

FuzzDistillML and FuzzDistillWeb rely on FuzzDistillCC to provide extracted program features as shown in Figure-\ref{fig:fuzzdistill_flowchart}.

The following sections discuss about each component and their design choices in detail -

\section{FuzzDistillCC : Feature Extraction}

\subsection{Compilation Insights}

The concept of leveraging compiler information to enhance fuzzing has been a longstanding practice within the fuzzing community, particularly for complex projects.

The fundamental principle remains consistent:
We want to fuzz of the binary $\rightarrow$ which subsequently originates from a compiler $\rightarrow$ This compiler possesses extensive knowledge about the program it compiled $\Rightarrow$ we can harness this information to gain enhanced insights into the fuzz surface.

These insights can be leveraged to guide fuzz testing efforts by identifying high-priority areas of the codebase that are more likely to harbor vulnerabilities.

The most common use of compilers in fuzzing is to instrument code for coverage information. Code coverage metrics, such as line, branch, or function coverage, can help identify areas of the codebase that are rarely executed or have limited testing exposure.

Several additional types of compiler insights can be particularly beneficial for fuzzing target selection, for example:

\begin{itemize}
\item Function Call Graphs: Compilers generate function call graphs, which depict the calling relationships among functions within the codebase. These graphs facilitate the identification of intricate function interactions, potential data flows, etc.

\item Data Flow Dependencies: By analyzing data flow dependencies, compilers can provide insights into how variables are used and propagated throughout the codebase. This information can be used to identify sensitive data handling routines, potential injection points, and areas where data validation is crucial.

\item Control Flow Graphs: Control flow graphs represent the flow of execution through the codebase, including conditional statements, loops, and function calls. These graphs can help identify complex control flows, potential error handling paths, and areas with high cyclomatic complexity.

\item Type System Information: By analyzing type system information, compilers can provide insights into data types, object relationships, and potential casting issues, which can be used to identify areas with high risk of type-related vulnerabilities.
\end{itemize}

In this project, I use two compiler passes to extract information from Basic Blocks and Functions.

Basic blocks represent the smallest unit of code that is free of control flow branches (other than the entry and exit points).

A function is a collection of logically connected basic blocks.

In LLVM, the compiler framework utilized in this project, programs are structured into modules. A module comprises functions, global variables, and symbol table entries. These modules can be combined using the LLVM linker, which merges function and global variable definitions, resolves forward declarations, and merges symbol table entries. \cite{noauthor_llvm_nodate}

\begin{figure}
    \centering
    \includegraphics[width=\linewidth]{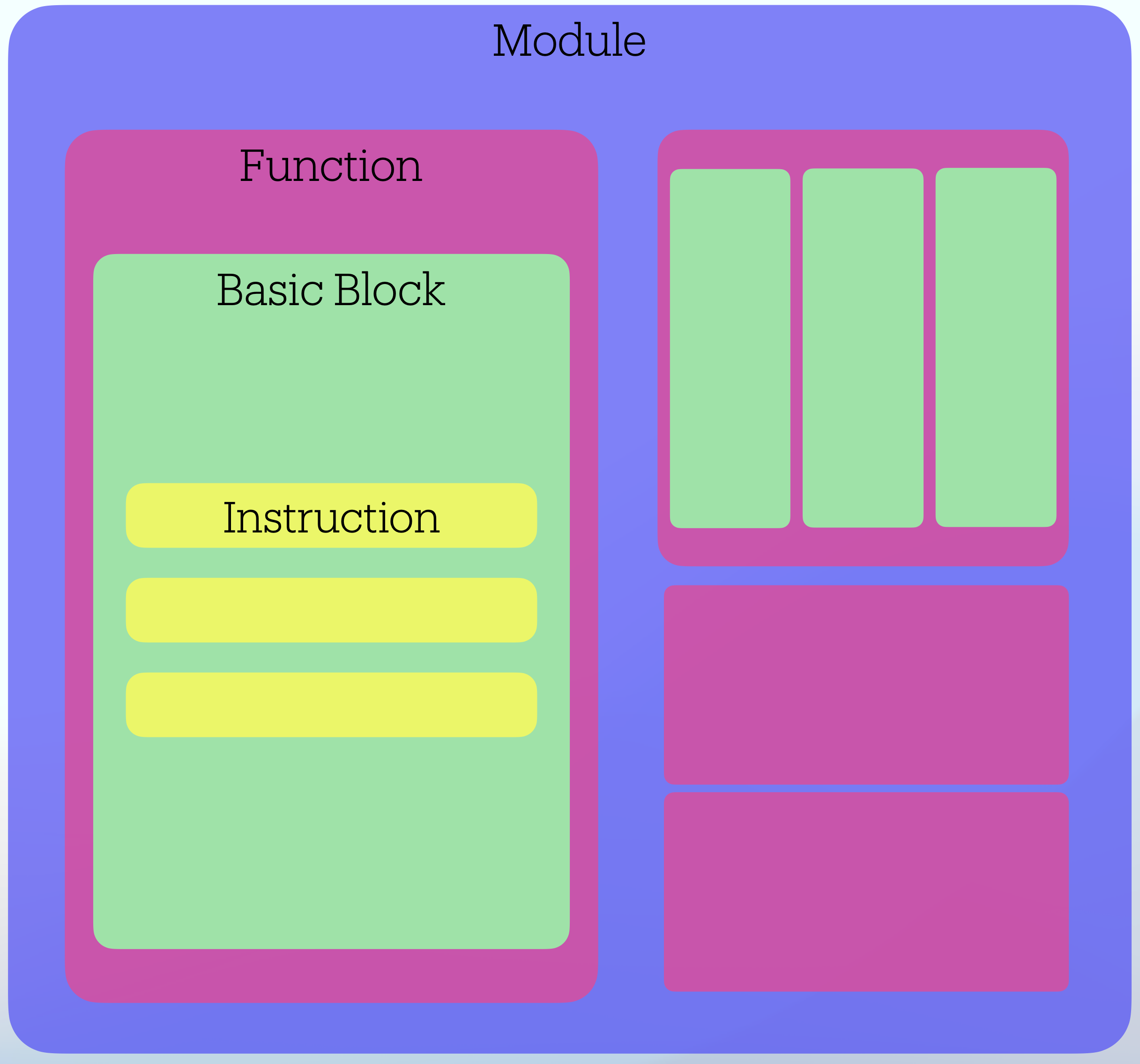}
    \caption{Structure of Module in LLVM}
    \label{fig:llvmstrmodule}
\end{figure}

A LLVM module follows a structured format, as outlined in Figure-\ref{fig:llvmstrmodule}.

\subsection{Basic Block Features}

FuzzDistillCC extracts the following features from each basic block:

\begin{enumerate}
\item \textbf{Block ID:} Unique ID associated to the block
\item \textbf{Block Name:} Name of the block with following structure - \lstinline{BB_<block #>_<demangled parent function>}
\item \textbf{Instructions:} Number of intermediate instructions in the block.
\item \textbf{In-degree:} Number of paths coming to the block.
\item \textbf{Out-degree:} Number of paths out of the block.
\item \textbf{Static  Allocations:} Number of static var allocations.
\item \textbf{Dynamic Allocations:} Number of dynamic var. allocations.
\item \textbf{MemOps:} Number of C/C++ memory operations (language dependent feature)
\item \textbf{CondBranches:} (Conditional Branches): The number of conditional branches (e.g., if-else statements) in the basic block.
\item \textbf{UnCondBranches:} (Unconditional Branches): The number of unconditional branches (e.g., jumps, returns) in the basic block.
\item \textbf{DirectCalls:} Number of direct function calls.
\item \textbf{InDirectCalls:}Number of in-direct function calls (call to pointers)
\item \textbf{VULNERABLE:} Training only label. 0 = safe, 1 = vulnerable. 
\end{enumerate}

Conditional branches (CondBranches) and unconditional branches (UnCondBranches) primarily serve as sanity checks and do not significantly impact the categorization of basic blocks (it might actually harm the accuracy). Let’s analyze the possible values of N (number of conditional branches) and M (number of unconditional branches).

A basic block can contain at most one conditional branch. A conditional branch is typically used to terminate the block and transfer control to another location within the code. If there were multiple conditional branches, they would need to be combined into a single decision point using logical operators, which would not increase the count of separate conditional branches. 
 \begin{equation}
 \therefore N \in \{0, 1\}
 \label{eq:niszo}
 \end{equation}
 where $N$ is either $0$ (no conditional branch) or $1$ (one conditional branch).
Similarly, a basic block can have at most one unconditional branch. An unconditional branch is typically used to exit the block and jump to another location in the code. If there were multiple unconditional branches, they would be redundant, as only one of them would be executed.
\begin{equation}
\therefore M \in \{0, 1\}
\label{eq:miszo}
\end{equation}
where $M$ is either $0$ (no unconditional branch) or $1$ (one unconditional branch).

If a basic block contains a conditional branch ($N = 1$), it is not possible to have an unconditional branch ($M = 0$), as the control flow would be determined solely by the conditional branch. Conversely, if a basic block includes an unconditional branch ($M = 1$), it is not feasible to have a conditional branch ($N = 0$), as the unconditional branch would override any conditional decision.
Logically - 
\begin{equation}
    N \times M = 0
    \label{eq:mniszo}
\end{equation}
\begin{equation}
(N = 1) \Rightarrow (M = 0)
\end{equation}
\begin{equation}
(M = 1) \Rightarrow  (N = 0)
\end{equation}

In essence, by eq-\ref{eq:niszo}, eq-\ref{eq:miszo}, and eq-\ref{eq:mniszo} only one of $N$ or $M$ can have the value of $1$ at any given time. If $N$ is set to $1$, $M$ must be set to $0$, and vice versa.

We can use this relationship to check the functionality of our compiler passes and sanity of our training dataset.

\subsection{Function Features}

\begin{enumerate}
\item \textbf{Function ID:} Unique \lstinline{unsigned long}
\item \textbf{Function Name:} Function's name (C++ function names are demangled using \lstinline{cxxabi})
\item \textbf{Instructions:} Number of intermediate instructions in the function.
\item \textbf{BBs:} Number of Basic Blocks in the function, (indication of control flow simplicity, lower the number, simpler the control flow.)
\item \textbf{In-degree:} Number of paths coming to the function.
\item \textbf{Out-degree:} Number of paths out of the function.
\item \textbf{Num Loops:} Number of loops in the function.
\item \textbf{Static  Allocations:} Number of static var allocations.
\item \textbf{Dynamic Allocations:} Number of dynamic var. allocations.
\item \textbf{MemOps:} Number of C/C++ memory operations (language dependent feature)
\item \textbf{CondBranches:} Number of conditional branches (if-else, switch, etc.)
\item \textbf{UnCondBranches:} Number of unconditional branches (calls, jumps, etc.)
\item \textbf{DirectCalls:} Number of direct function calls.
\item \textbf{InDirectCalls:}Number of in-direct function calls (call to pointers)
\item \textbf{VULNERABLE:} Training only label. 0 = safe, 1 = vulnerable. 
\end{enumerate}

\section{Dataset Generation and Feature Selection}

The final Basic Block and Function training data is presented in Semicolon-Separated Values (SSV) format.

For generating training data, I utilized NIST Juliet C/C++ 1.3 \cite{noauthor_juliet_nodate}, a comprehensive collection of test cases in the C/C++ programming language. This resource is structured into 118 distinct Common Weakness Enumeration (CWE) categories, providing a diverse range of examples for training purposes. 

Generating the final SSV dataset presents a challenge due to Juliet’s modular, per-CWE test cases. Compiling each test case using FuzzDistillCC results in a small SSV for that specific test case.
To address this issue, I compile all cases using FuzzDistillCC and generate per-example SSV data without headers. Subsequently, I traverse the directories and concatenate individual SSV files into a single final SSV file, and prepend the associated data header.

\subsection{Feature Selection}
During training I droped the features in \lstinline{TARGET}, \lstinline{UNIMPORTANT} and \lstinline{EXPLICIT_EXCLUDE} categories:
\begin{lstlisting}[language=python, caption=function feature categories]
TARGET_FEATURE = ["VULNERABLE"]
UNIMPORTANT_FEATURES = [
    "FunctionID",
    "FunctionName"
    ]
CPP_MEMORY_FEATURES = ["MemOps"]
EXPLICIT_EXCLUDE_FEATURES =[
    "InDirectCalls"
    ]
\end{lstlisting}

We can also drop \lstinline{CPP_MEMORY_FEATURES}, as it doesn't seem to have a significant impact on decision as discussed later in Section \ref{subsec:bestfeatures}.

For Basic blocks, I follow similar setup with following feature categories - 

\begin{lstlisting}[language=python, caption=block feature categories]
BB_TARGET_FEATURE = ["VULNERABLE"]
BB_UNIMPORTANT_FEATURES = [
    "BlockID",
    "BlockName"
    ]
BB_CPP_MEMORY_FEATURES = ["MemOps"]
BB_EXPLICIT_EXCLUDE_FEATURES=[
    "CondBranches",
    "InDirectCalls",
    "UnCondBranches",
    "MemOps"
    ]

\end{lstlisting}

\section{FuzzDistillML: Model Training}

The nature of this problem falls under the category of ``binary classification'', which involves identifying features that either represent a safe or vulnerable block or function.
I explored various avenues for model training, including Instance-Based Models, Tree-Based Models, Linear Models, Probabilistic Models, and Neural Networks.

At the end I decided to integrate Neural Network model with the web front-end, but there was no significant advantage of this decision.

The following subsections discuss both approaches in detail.

The following Notebooks in \lstinline{Training/} directory of FuzzDistillML contains tests and drafts of all models I tried
\begin{lstlisting}
testmultiplemodelsFN.ipynb
testmultiplemodelsBB.ipynb
tensorflowTrainerBB.ipynb
tensorflowTrainerFN.ipynb
\end{lstlisting}
and \lstinline{Training/final_training_scripts/} contains selected  final methods.

While experimenting with various algorithms, including:
\begin{itemize}
    \item Logistic Regression
    \item Random Forest
    \item Support Vector Machines (SVM)
    \item Gradient Boosting
    \item K-Nearest Neighbors (KNN)
    \item Naive Bayes
    \item Decision Trees
    \item AdaBoost
\end{itemize}

Extreme Gradient Boosting (XGBoost) and Neural Network demonstrated the most favorable outcome.

\subsection{Extreme Gradient Boosting}
I used a eXtreme Gradient Boosting (XGBoost) classifier to predict the probability of classes (binary) in my dataset. The algorithm is a widely adopted and effective method for handling intricate datasets with non-linear relationships.\cite{noauthor_dmlcxgboost_2024}

The XGBoost model was trained on my dataset with the following hyperparameters:
\begin{lstlisting}[language=python, caption=XGBoost Model Definition]
model = xgb.XGBClassifier(
  objective='binary:logistic',
  eval_metric='logloss',
  random_state=40,
  colsample_bytree=0.8,
  learning_rate=0.05,
  max_depth=10,
  n_estimators=400,
  subsample=0.8
)
\end{lstlisting}
These hyperparameters were selected through a combination of grid search and cross-validation to optimize the model's performance.

\begin{figure}
    \centering
    \includegraphics[width=\linewidth]{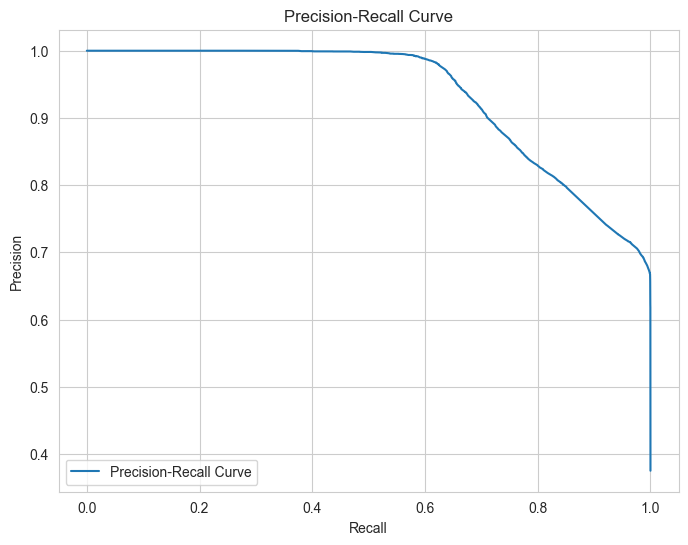}
    \caption{XGBoost Precision-Recall curve}
    \label{fig:xgrecallcrv}
\end{figure}

\begin{figure}
    \centering
    \includegraphics[width=\linewidth]{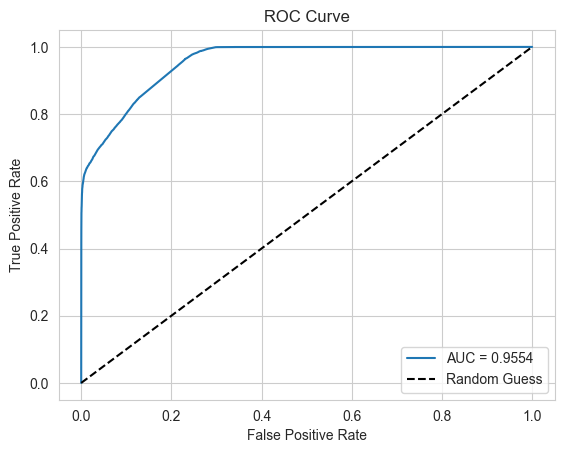}
    \caption{XGBoost ROC curve}
    \label{fig:xgroc}
\end{figure}

The trained XGBoost model demonstrated an accuracy of 86.31\% on  test dataset. In addition to accuracy, I evaluated the model’s performance using various metrics, including precision (82.75\%), recall (80.24\%) (Figure-\ref{fig:xgrecallcrv}), F1 score (81.47\%), and area under the receiver operating characteristic curve (AUC-ROC) (95.54\%) (Figure-\ref{fig:xgroc}). 

\begin{figure}
    \centering
    \includegraphics[width=\linewidth]{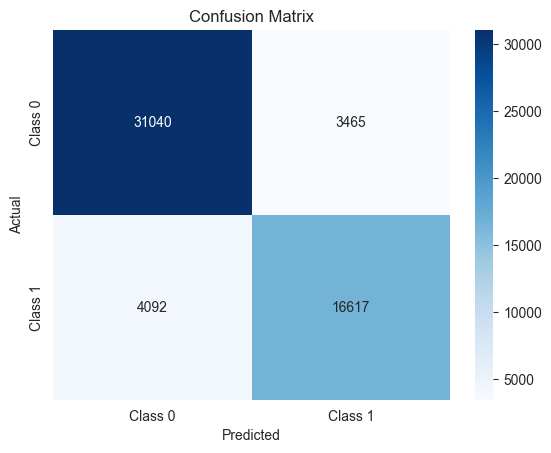}
    \caption{XGBoost Confusion Matrix}
    \label{fig:xgconfmats}
\end{figure}

The confusion matrix in Figure-\ref{fig:xgconfmats} offers a comprehensive analysis of the model’s performance, indicating 31,040 true negatives, 3,465 false positives, 4,092 false negatives, and 16,617 true positives. The significant number of true positives highlights the model’s effectiveness in identifying vulnerable samples.

\begin{figure}
    \centering
    \includegraphics[width=\linewidth]{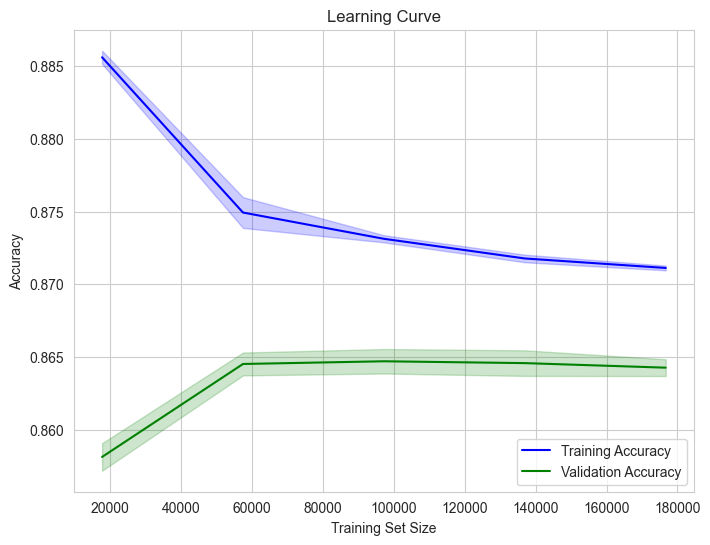}
    \caption{XGBoost Learning Curve}
    \label{fig:xglearningcurve}
\end{figure}

Figure-\ref{fig:xglearningcurve} shows the learning curve for the model.
Training Accuracy (Blue Curve) starts at a high value ($\approx$0.885) with a small training set (expected overfitting with limited data). Subsequently, it decreases as more data is incorporated. As the dataset size expands, the model exhibits improved generalization, resulting in a lower training accuracy. However, this trend stabilizes with a sufficiently large training set.
Validation Accuracy (Green Curve) starts at a lower value ($\approx$0.855) and subsequently increases with dataset growth. It reaches a peak and stabilizes around 0.865.
Increasing the training data enhances the model’s generalization capabilities to unseen data.
The initial disparity between training and validation accuracy indicates overfitting, which narrows with larger datasets.
The shaded regions encompass the variability.
Narrowing the confidence intervals for both curves suggests more stable model performance with larger datasets.
Overall, the model demonstrates effective generalization with increased training data, as evidenced by the narrowing training-validation accuracy gap.

These results suggest that  XGBoost model effectively distinguishes between vulnerable and non-vulnerable samples.

\subsection{Deep Neural Network}

\begin{figure}
    \centering
    \includegraphics[width=0.3\linewidth]{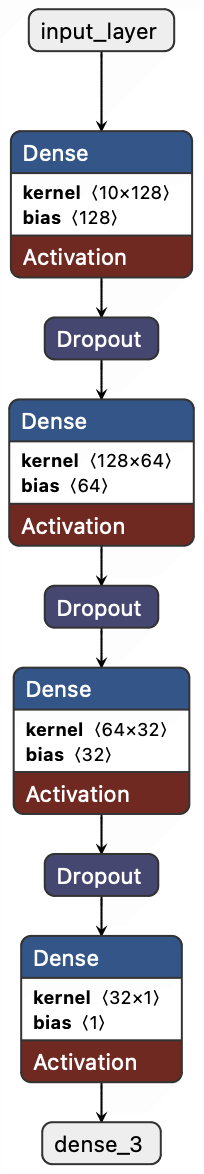}
    \caption{Structure of DNN}
    \label{fig:dnnnetworkgraph}
\end{figure}

For my Deep Neural Network (DNN) approach, I utilized the Keras Sequential API within the TensorFlow framework. The architecture comprises four fully connected (dense) layers, each employing the ReLU activation function. Subsequently, dropout layers are incorporated to mitigate the risk of overfitting. The model’s structure comprises 128 units in the initial hidden layer, 64 units in the second hidden layer, and 32 units in the third hidden layer, all with ReLU activation. The output layer comprises a single unit with the sigmoid activation function, suitable for binary classification scenarios (vulnerable or safe). (Figure-\ref{fig:dnnnetworkgraph})

The model was compiled using the Adam optimizer, with a learning rate of 0.001 and the binary cross-entropy loss function as the loss criterion. 

To mitigate the risk of overfitting, early stopping was employed with a patience setting of 5 epochs. This strategy enabled the restoration of the weights associated with the lowest validation loss, thereby ensuring the most effective model.

The model was trained for a maximum of 30 epochs using a batch size of 32 and the Adam optimizer. A 20\% stratified test split was employed for testing and validation purposes.

The model achieved an overall accuracy of 86\%, with a high precision and recall for Class 0 (minority class) at 88\% and 90\% respectively. However, performance metrics for Class 1 (majority class) were slightly lower at 82\% precision and 80\% recall. Despite this, the F1-scores and macro-averaged precision and recall indicate a robust model that balances performance across both classes.

\begin{figure}
    \centering
    \includegraphics[width=\linewidth]{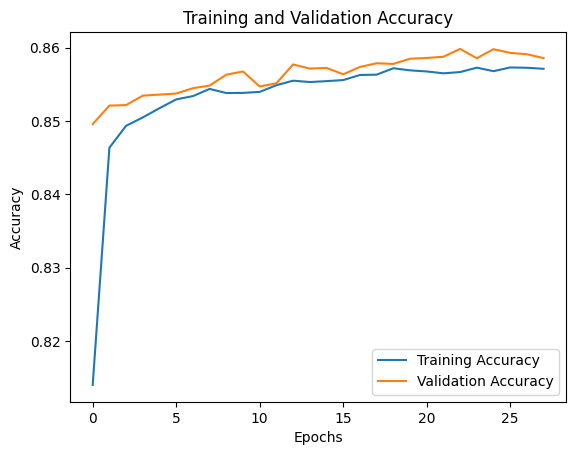}
    \caption{Training and validation accuracy}
    \label{fig:dnnaccuracy}
\end{figure}

\begin{figure}
    \centering
    \includegraphics[width=\linewidth]{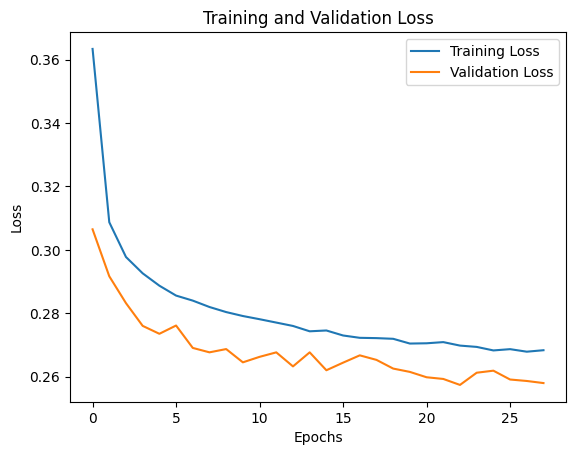}
    \caption{Training and validation loss}
    \label{fig:dnnloss}
\end{figure}

\begin{figure}
    \centering
    \includegraphics[width=\linewidth]{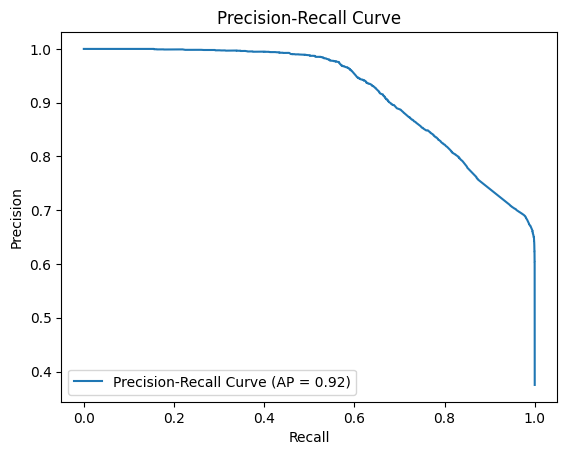}
    \caption{Precision-Recall Curve}
    \label{fig:dnnrecall}
\end{figure}

\begin{figure}
    \centering
    \includegraphics[width=\linewidth]{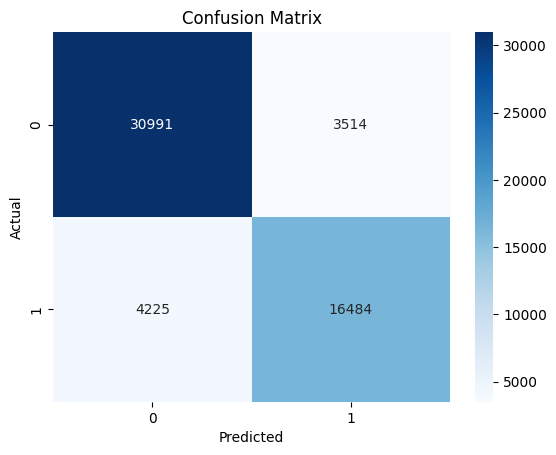}
    \caption{DNN Confusion Matrix}
    \label{fig:dnnconfmat}
\end{figure}

Figure \ref{fig:dnnloss} presents the training and validation loss over the epochs. The model exhibits a consistent decrease in training loss while the validation loss plateaus after a few epochs, indicating that the EarlyStopping mechanism effectively prevents overfitting.
Figure \ref{fig:dnnaccuracy} illustrates the training and validation accuracy. Similar to the loss curves, the model shows increasing training accuracy with stable validation accuracy, suggesting a well-performing model on both seen and unseen data.
Figure \ref{fig:dnnrecall} presents the recall for both classes over epochs. The recall for Class 0 is consistently high, indicating strong performance in identifying this class, while Class 1 has a slightly lower but still acceptable recall.
Figure \ref{fig:dnnconfmat} presents the confusion matrix, providing a detailed breakdown of true positives, false positives, true negatives, and false negatives for both classes.

\begin{figure}
    \centering
    \includegraphics[width=\linewidth]{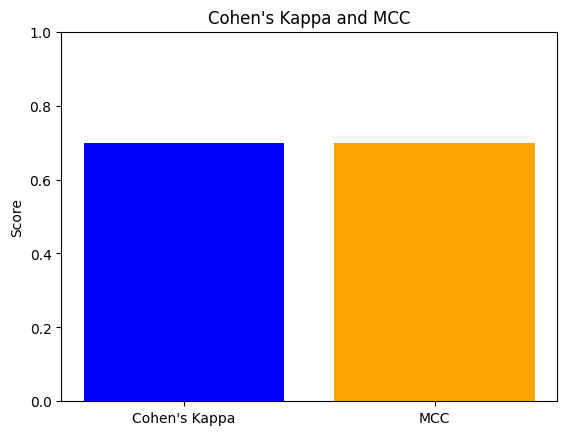}
    \caption{Matthews Correlation Coefficient and Cohen’s Kappa}
    \label{fig:dnnckmcc}
\end{figure}

Additionally, the Matthews Correlation Coefficient (MCC) of 0.6992 indicates a strong positive correlation between the predicted and actual classifications. Furthermore, 0.6989 Cohen’s Kappa suggests substantial agreement between the model’s predictions and the true labels. (Figure-\ref{fig:dnnckmcc})


\subsection{Best Features}
\label{subsec:bestfeatures}
\begin{figure}
    \centering
    \includegraphics[width=\linewidth]{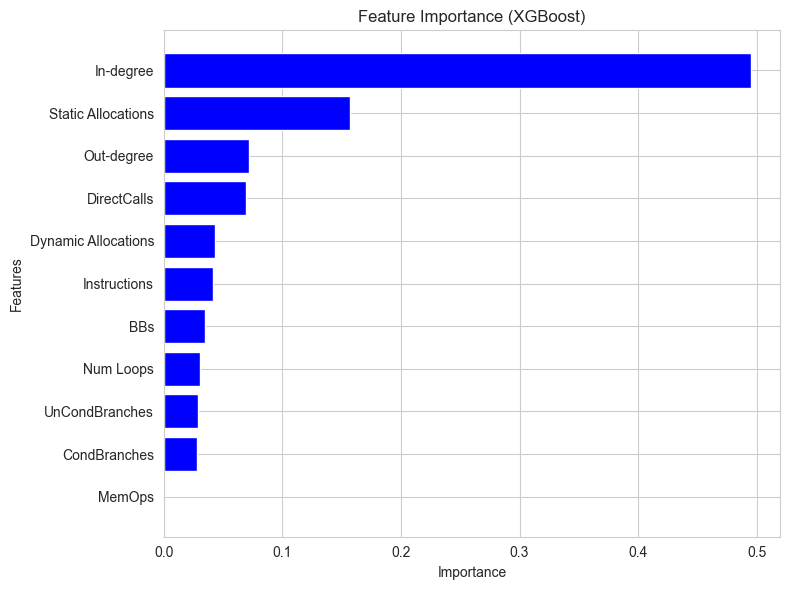}
    \caption{XGBoost Important Features}
    \label{fig:xgimpfeature}
\end{figure}

XGBoost provides feature importance scores. In my function feature dataset, as shown in Fig-\ref{fig:xgimpfeature}, the five most influential features (by weight) were In-degree, Static Allocations, Out-degree, Direct Calls, and Dynamic Allocations.

\begin{figure}
    \centering
    \includegraphics[width=\linewidth]{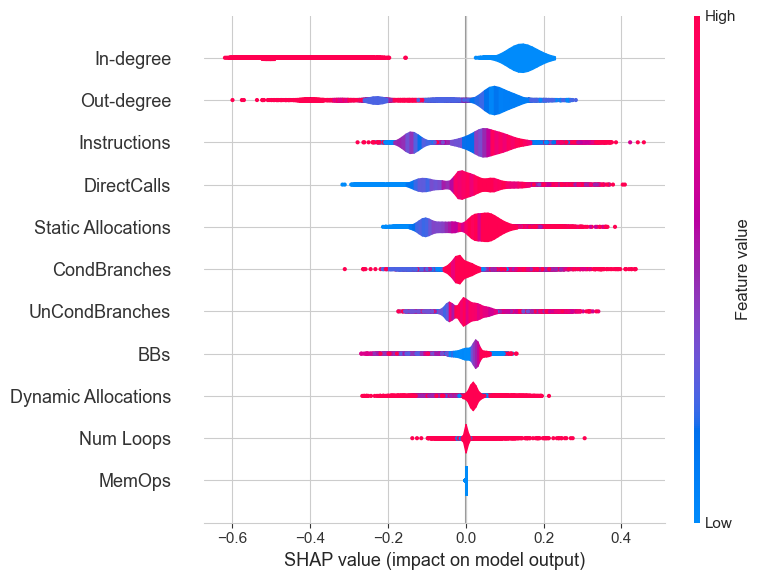}
    \caption{DNN Important Features}
    \label{fig:dnnfeatures}
\end{figure}

For DNN I used shapley additive explanation (SHAP)\cite{noauthor_welcome_nodate, noauthor_introduction_nodate}.
Figure \ref{fig:dnnfeatures} illustrates the impact of features on the output of the neural network model.
The top five features that significantly influence the model’s output are: In-degree, Out-degree, Instructions, Direct Calls, Static Allocations

\subsection{Hyper parameter Tuning}
To get the best configuration possible, I used \lstinline{optuna}, a hyperparameter optimization framework in python \cite{akiba_optuna_2019} to optimize DNN. The code is available at \lstinline{Training/tuneHyperTFforFN.ipynb}.

For finding optimal parameters for XGBoost, I used \lstinline{GridSearchCV} from \lstinline{sklearn.model_selection}.

\section{FuzzDistillWeb: Front-end}

FuzzDistillWeb implements a Flask web app that processes CSV files with function data and predicts vulnerabilities using trained models. It handles file uploads, generates predictions with confidence scores, and filters/categorizes vulnerabilities. The app generates visualizations (bar charts, pie charts) to summarize prediction metrics. These visualizations are base64-encoded for web page integration. It also provides HTML tables for detailed listings of all identified vulnerable functions.

To optimize performance and resource utilization, I have implemented a in-memory local cache that stores previously processed results. This enables efficient retrieval of results for the same feature sets, thereby reducing the need for repetitive processing and associated time and resource consumption.

The cache does not automatically clear. The application exposes two developer APIs: one to clear a record from the cache and another to purge the entire cache. This action will reprocess the record on the next query.

\subsection{API Documentation}

This subsection provides details about the API endpoints exposed by the Flask application. These endpoints allow interaction with the underlying machine learning models and cache management system.

\subsubsection{Prediction APIs}

\begin{itemize}
    \item \textbf{\texttt{POST /api/high-conf-list}}
    \begin{itemize}
        \item \textbf{Description:} Returns a list of high-confidence predictions.
        \item \textbf{Request Parameters:}
        \begin{itemize}
            \item \texttt{file}: A CSV file containing the data to be processed.
            \item \texttt{modelselect}: The model to use for predictions (\texttt{dnnfn} or \texttt{dnnbb}).
        \end{itemize}
    \end{itemize}

    \item \textbf{\texttt{POST /api/sure-list}}
    \begin{itemize}
        \item \textbf{Description:} Returns a list of predictions with 100\% confidence.
        \item \textbf{Request Parameters:}
        \begin{itemize}
            \item \texttt{file}: A CSV file containing the data to be processed.
            \item \texttt{modelselect}: The model to use for predictions (\texttt{dnnfn} or \texttt{dnnbb}).
        \end{itemize}
    \end{itemize}

    \item \textbf{\texttt{POST /api/all-list}}
    \begin{itemize}
        \item \textbf{Description:} Returns all vulnerable predictions.
        \item \textbf{Request Parameters:}
        \begin{itemize}
            \item \texttt{file}: A CSV file containing the data to be processed.
            \item \texttt{modelselect}: The model to use for predictions (\texttt{dnnfn} or \texttt{dnnbb}).
        \end{itemize}
    \end{itemize}
\end{itemize}

\subsubsection{Cache Management APIs}

\begin{itemize}
    \item \textbf{\texttt{GET /api/clear-cache-record}}
    \begin{itemize}
        \item \textbf{Description:} Clears the cache for a specific file.
        \item \textbf{Query Parameters:}
        \begin{itemize}
            \item \texttt{hash}: The sha256 hash of the file to be cleared.
        \end{itemize}
    \end{itemize}

    \item \textbf{\texttt{POST/GET /api/clear-cache}}
    \begin{itemize}
        \item \textbf{Description:} Clears the entire cache.
    \end{itemize}
\end{itemize}

\section{Future Work}
This project serves as a preliminary demonstration of the concept’s feasibility. While it may not be the pioneering work of its kind, I provide the foundational infrastructure and a compelling impetus for reproducibility of this approach. Furthermore, I present the scope of potential future collaborations and numerous avenues for enhancement.

Future endeavors could prioritize optimizing hyperparameters further, investigating alternative machine learning algorithms, or integrating supplementary features into the dataset to augment the model’s performance.
The study employed a single dataset, potentially limiting its generalizability to diverse classification scenarios.
The model was trained with a predetermined set of hyperparameters, which may not be optimal for all applications.
The application of early stopping and dropout regularization may not be sufficient to mitigate overfitting in all instances.

\section{Reproducibility}

Parent Repository:
\begin{itemize}
    \item \url{https://github.com/Saket-Upadhyay/FuzzDistill}
\end{itemize}

Component Repositories:
\begin{itemize}
    \item \url{https://github.com/Saket-Upadhyay/FuzzDistillCC}
    \item \url{https://github.com/Saket-Upadhyay/FuzzDistillML}
    \item \url{https://github.com/Saket-Upadhyay/FuzzDistillWeb}
\end{itemize}

\section{Acknowledgements}
This project was facilitated by the opportunities provided in the graduate class ``CS 6501: Software Security Testing'', led by Dr. Jack Davidson, Dr. Jason Hiser, and Dr. Anh Nguyen-Tuong at the University of Virginia.
I would like to express my sincere gratitude to all the instructors for creating an environment that fostered exploration and innovation, enabling me to delve into the intersection of fuzz testing and machine learning. The guidance and support received throughout the course were instrumental in shaping this project.

\printbibliography

@inproceedings{zhu_regression_2021,
	location = {New York, {NY}, {USA}},
	title = {Regression Greybox Fuzzing},
	isbn = {978-1-4503-8454-4},
	url = {https://doi.org/10.1145/3460120.3484596},
	doi = {10.1145/3460120.3484596},
	series = {{CCS} '21},
	pages = {2169--2182},
	booktitle = {Proceedings of the 2021 {ACM} {SIGSAC} Conference on Computer and Communications Security},
	publisher = {Association for Computing Machinery},
	author = {Zhu, Xiaogang and Böhme, Marcel},
	urldate = {2024-12-08},
	date = {2021-11-13},
}

@inproceedings{weissberg_sok_2024,
	location = {Singapore Singapore},
	title = {{SoK}: Where to Fuzz? Assessing Target Selection Methods in Directed Fuzzing},
	url = {https://dl.acm.org/doi/10.1145/3634737.3661141},
	doi = {10.1145/3634737.3661141},
	shorttitle = {{SoK}},
	eventtitle = {{ASIA} {CCS} '24: 19th {ACM} Asia Conference on Computer and Communications Security},
	pages = {1539--1553},
	booktitle = {Proceedings of the 19th {ACM} Asia Conference on Computer and Communications Security},
	publisher = {{ACM}},
	author = {Weissberg, Felix and Möller, Jonas and Ganz, Tom and Imgrund, Erik and Pirch, Lukas and Seidel, Lukas and Schloegel, Moritz and Eisenhofer, Thorsten and Rieck, Konrad},
	urldate = {2024-12-09},
	date = {2024-07},
	langid = {english},
}

@inproceedings{haller_dowsing_2013,
	location = {{USA}},
	title = {Dowsing for overflows: a guided fuzzer to find buffer boundary violations},
	isbn = {978-1-931971-03-4},
	series = {{SEC}'13},
	shorttitle = {Dowsing for overflows},
	pages = {49--64},
	booktitle = {Proceedings of the 22nd {USENIX} conference on Security},
	publisher = {{USENIX} Association},
	author = {Haller, Istvan and Slowinska, Asia and Neugschwandtner, Matthias and Bos, Herbert},
	urldate = {2024-12-08},
	date = {2013-08-14},
}

@inproceedings{osterlund_parmesan_2020,
	title = {\{{ParmeSan}\}: Sanitizer-guided Greybox Fuzzing},
	isbn = {978-1-939133-17-5},
	url = {https://www.usenix.org/conference/usenixsecurity20/presentation/osterlund},
	shorttitle = {\{{ParmeSan}\}},
	eventtitle = {29th {USENIX} Security Symposium ({USENIX} Security 20)},
	pages = {2289--2306},
	author = {Österlund, Sebastian and Razavi, Kaveh and Bos, Herbert and Giuffrida, Cristiano},
	urldate = {2024-12-09},
	date = {2020},
	langid = {english},
}

@online{noauthor_introduction_nodate,
	title = {An introduction to explainable {AI} with Shapley values — {SHAP} latest documentation},
	url = {https://shap.readthedocs.io/en/latest/example_notebooks/overviews/An%20introduction%20to%20explainable%20AI%20with%20Shapley%20values.html},
	urldate = {2024-12-08},
}

@online{noauthor_welcome_nodate,
	title = {Welcome to the {SHAP} documentation — {SHAP} latest documentation},
	url = {https://shap.readthedocs.io/en/latest/index.html},
	urldate = {2024-12-08},
}

@software{noauthor_dmlcxgboost_2024,
	title = {dmlc/xgboost},
	rights = {Apache-2.0},
	url = {https://github.com/dmlc/xgboost},
	publisher = {Distributed (Deep) Machine Learning Community},
	urldate = {2024-12-07},
	date = {2024-12-07},
	note = {original-date: 2014-02-06T17:28:03Z},
}

@online{noauthor_saket-upadhyayfuzzdistillweb_nodate,
	title = {Saket-Upadhyay/{FuzzDistillWeb}},
	url = {https://github.com/Saket-Upadhyay/FuzzDistillWeb},
	titleaddon = {{GitHub}},
	urldate = {2024-12-07},
	langid = {english},
}

@online{noauthor_saket-upadhyayfuzzdistillml_nodate,
	title = {Saket-Upadhyay/{FuzzDistillML}: Traning data from {FuzzDistillCC}},
	url = {https://github.com/Saket-Upadhyay/FuzzDistillML},
	shorttitle = {Saket-Upadhyay/{FuzzDistillML}},
	titleaddon = {{GitHub}},
	urldate = {2024-12-07},
	langid = {english},
}

@online{noauthor_saket-upadhyayfuzzdistillcc_nodate,
	title = {Saket-Upadhyay/{FuzzDistillCC}: {FuzzDistill} Compiler Component- Feature Extraction Compiler Passes},
	url = {https://github.com/Saket-Upadhyay/FuzzDistillCC},
	shorttitle = {Saket-Upadhyay/{FuzzDistillCC}},
	titleaddon = {{GitHub}},
	urldate = {2024-12-07},
	langid = {english},
}

@software{akiba_optuna_2019,
	title = {Optuna: A next-generation hyperparameter optimization framework},
	rights = {{MIT}},
	url = {https://github.com/optuna/optuna},
	shorttitle = {Optuna},
	author = {Akiba, Takuya and Sano, Shotaro and Yanase, Toshihiko and Ohta, Takeru and Koyama, Masanori},
	urldate = {2024-12-07},
	date = {2019},
	doi = {10.1145/3292500.3330701},
	note = {Pages: 2623–2631
Publication Title: Proceedings of the 25th {ACM} {SIGKDD} international conference on knowledge discovery \& data mining
original-date: 2018-02-21T06:12:56Z},
}

@online{noauthor_juliet_nodate,
	title = {Juliet C/C++ 1.3},
	url = {https://samate.nist.gov/SARD},
	titleaddon = {{NIST} Software Assurance Reference Dataset},
	urldate = {2024-12-07},
	langid = {english},
}

@online{noauthor_llvm_nodate,
	title = {{LLVM} Language Reference Manual — {LLVM} 20.0.0git documentation},
	url = {https://llvm.org/docs/LangRef.html#module-structure},
	urldate = {2024-12-07},
}
\end{document}